\documentclass[journal,twoside,web]{ieeecolor}

\usepackage{etoolbox}
\makeatletter
\@ifundefined{color@begingroup}%
{\let\color@begingroup\relax
\let\color@endgroup\relax}{}%
\def\fix@ieeecolor@hbox#1{%
\hbox{\color@begingroup#1\color@endgroup}}
\patchcmd\@makecaption{\hbox}{\fix@ieeecolor@hbox}{}{\FAILED}
\patchcmd\@makecaption{\hbox}{\fix@ieeecolor@hbox}{}{\FAILED}

\usepackage{tmi}
\usepackage{cite}
\usepackage{amsmath,amssymb,amsfonts}
\usepackage{algorithmic}
\usepackage{graphicx}
\usepackage{textcomp}
\usepackage{booktabs}
\usepackage{url}
\usepackage{multirow,multicol}
\usepackage{pifont}       % \ding{xx}
\usepackage{colortbl}

\def\etal{{\emph{et al.~}}}
\newtheorem{assump}{Assumption}

\def\BibTeX{{\rm B\kern-.05em{\sc i\kern-.025em b}\kern-.08em
    T\kern-.1667em\lower.7ex\hbox{E}\kern-.125emX}}
\begin{document}
% \title{Contour-weighted loss function for imbalanced image segmentation with partial decoder attention network}
\title{Partial Decoder Attention Network with Contour Weighted Loss Function for Data-Imbalance Medical Image Segmentation}

\author{Zhengyong Huang, Ning Jiang, Xingwen Sun, Lihua Zhang, Peng Chen, Jens Domke and Yao Sui%, \IEEEmembership{Member, IEEE}
\thanks{Research reported in this study was supported by the Major Program of the National Natural Science Foundation of China, under Award Numbers 62394310 and 62394312.}
\thanks{Zhengyong Huang and Ning Jiang are with the Institute of Medical Technology, Peking University Health Science Center, Peking University, Beijing, China, and with the National Institute of Health Data Science, Peking University, Beijing, China.}
\thanks{Xingwen Sun and Lihua Zhang are with the Department of Radiology, Peking University Third Hospital, Beijing, China.}
\thanks{Peng Chen and Jens Domke are with RIKEN Center for Computational Science (R-CCS), Kobe, Japan.}
\thanks{Yao Sui is with the National Institute of Health Data Science, Peking University, the Institute of Medical Technology, Peking University Health Science Center, and the Institute for Artificial Intelligence, Peking University, Beijing, China.}
\thanks{Corresponding author: Yao Sui (yaosui@pku.edu.cn).}
}
\maketitle

\begin{abstract}
% background
Image segmentation is pivotal in medical image analysis, facilitating clinical diagnosis, treatment planning, and disease evaluation. 
%Various segmentation techniques have been implemented for medical images. 
In recent years, deep learning has significantly advanced automatic segmentation methodologies by providing superior modeling capability for complex structures and fine-grained anatomical regions. However, medical images often suffer from data imbalance issues, such as large volume disparities among organs or tissues, and uneven sample distributions across different anatomical structures. This imbalance tends to bias the model toward larger organs or more frequently represented structures, while overlooking smaller or less represented structures, thereby affecting the segmentation accuracy and robustness.
%
% methods
To address these challenges, we proposed a novel contour-weighted segmentation approach, which improves the model's capability to represent small and underrepresented structures. We developed PDANet, a lightweight and efficient segmentation network based on a partial decoder mechanism. We evaluated our method using three prominent public datasets.
%
% results
The experimental results show that our methodology excelled in three distinct tasks: segmenting multiple abdominal organs, brain tumors, and pelvic bone fragments with injuries. It consistently outperformed nine state-of-the-art methods while reducing the complexity of the model. Moreover, the proposed contour-weighted strategy improved segmentation for other comparison methods across the three datasets, yielding average enhancements in Dice scores of 2.39\%, 1.44\%, and 3.88\%, respectively.
%
%conclusions
The results demonstrate that our contour-weighted segmentation method surpassed current leading approaches in both accuracy and robustness. As a model-independent strategy, it can seamlessly fit various segmentation frameworks, enhancing their performance. This flexibility highlighted its practical importance and potential for broad use in medical image analysis.

\end{abstract}


\section{Introduction}
\label{sec: Introduction}
\IEEEPARstart{A}{utomatic} delineation of relevant organs and tissues is important for clinical tasks such as diagnosis, treatment planning, and disease evaluation \cite{huang2022isa}. Image segmentation has become the primary technique to achieve this goal, often acting as the foundational step for quantitative evaluations of anatomical structures \cite{hatamizadeh2022unetr,zhao2025uncertainty}. 
Consequently, the efficacy of image segmentation profoundly influences the quality of medical image analysis \cite{zhu2025merging}. 
% Despite significant advancements in image segmentation methodologies in recent years \cite{cao2022swin,lin2025boosting,chen2024transunet,zhang2024cqformer}, 
Despite significant advancements in image segmentation methodologies in recent years \cite{lin2025boosting,chen2024transunet,zhang2024cqformer}, 
there remain gaps between these techniques and their practical application scenarios \cite{chen2022recent}. A notable challenge is the imbalanced nature of data \cite{wang2023dhc,pan2023smile,guo2024imbalanced}, which frequently leads to incorrect segmentation results, such as false positives or negatives. 
Data imbalance is prevalent in medical image segmentation, arising from both volume disparities between local structures and the uneven distribution of the sample quantity between different anatomical structures \cite{wang2023dhc}. 
This issue is particularly pronounced in tasks such as abdominal multi-organ segmentation~\cite{wang2023dhc,pei2023pets,chen2024multi}, brain tumor segmentation~\cite{pei2023pets}, and pelvic bone fragments with injuries segmentation~\cite{LIUandYIBULAYIMU2025MEDIA,liu2023pelvic}, where uneven distributions of data can cause errors in identifying small or rare anatomical structures. 
The issue of data imbalance leads to model bias towards larger organs or more frequently represented structures, resulting in inadequate feature learning for smaller or less represented structures. Therefore, it is essential to address the data imbalance in practical applications.

With the recent success of deep learning in semantic segmentation, various deep learning-based methods have emerged for medical image segmentation \cite{liu2023clip, hu2023label,zhou2025efficient}. These methods usually involve carefully handcrafted deep neural networks to achieve an efficient feature representation for medical images. 
To improve segmentation efficacy, certain studies have focused on the formulation of computationally intensive networks with extensive parameter sets \cite{karimijafarbigloo2025medscale,li2024spatial} or the implementation of a variety of data augmentation techniques \cite{zhang2025unsupervised,li2023joint}. However, these methods often lack flexibility in real-world scenarios and are challenging to deploy in resource-constrained settings \cite{xu2025vsnet}. In contrast, optimizing a loss function can offer a lightweight, flexible, and model-independent solution that improves segmentation without appreciably increasing model complexity \cite{zhu2022compound}. This study is thus committed to advancing medical image segmentation by refining the loss function to improve accuracy and robustness while maintaining model simplicity.

Current methodologies for improving loss functions can be divided primarily into three categories: distribution-based, region-based, and compound loss functions \cite{zhu2022compound}. 
Distribution-based loss functions, such as cross-entropy loss and its enhanced variants 
\cite{wang2019symmetric,akil2020fully,caliva2019distance}, minimize the disparities between two probability distributions. While these losses effectively manage both false positives and negatives, they present significant challenges in scenarios with data imbalance. 
For example, cross-entropy loss tends to be dominated by abundant samples or large volumes and is sensitive to hard-to-classify regions, which together impair the model’s ability to accurately delineate small and underrepresented structures.

Numerous weighted cross-entropy methods have been proposed to address these issues. For example, weighted cross-entropy (WCE) assigns higher weights to small or underrepresented anatomical structures to compensate for imbalances. However, determining optimal weights is challenging, as manually set weights may not generalize well across varied datasets. Caliva \etal \cite{caliva2019distance} proposed a distance map-weighted cross-entropy loss (DWCE) to tackle data imbalance, reporting favorable results. This method allocates greater weights to pixels near boundaries, producing a weighted cross-entropy loss. Focal loss \cite{lin2017focal}, a reshaped cross-entropy loss function, is specifically designed to address data imbalance by reducing the weights of well-classified examples, thereby focusing training efforts on more challenging cases. However, these losses treat pixels independently, failing to capture spatial dependencies, thus diminishing their effectiveness for volumetric segmentation.

Region-based loss functions, such as Dice loss and its derivatives \cite{milletari2016v,sudre2017generalised}, strive to maximize the overlap between predicted and ground truth images. This characteristic inherently makes them robust in managing the volume disparities. Unlike cross-entropy-based losses, which process pixels independently, Dice loss assesses the local structure by computing set-based similarity, ensuring improved segmentation of small or less-represented targets. Additionally, it enhances boundary precision by emphasizing region-level consistency rather than individual pixel-wise accuracy. However, Dice loss is sensitive to small segmentation volume, where even minor misclassifications can cause significant gradient fluctuations, leading to instability during training \cite{huang2024contour}. 
Although Sudre \etal \cite{sudre2017generalised} proposed generalized Dice loss (GDL) to allocate weights based on target volume and thereby mitigate data imbalance of different segmentation targets, the issue of training stability persists.

To mitigate these challenges, researchers have explored combinations of different loss functions to take advantage of their complementary benefits \cite{guo2024randomness,li2023joint}. For instance, the Combo loss \cite{zhu2022compound} integrates Dice loss with cross-entropy loss, effectively bridging distribution-based and region-based approaches. Another common composite loss fuses Dice loss with focal loss. These combinations enhance pixel-level classification accuracy while preserving region-level structural consistency, leading to improved segmentation.  

Based on the above analysis, we proposed a novel compound loss function to address the prevalent data imbalance challenges in medical image segmentation, specifically those arising from volume disparities and sample distribution imbalances. Regardless of the type of imbalance, segmentation errors are often concentrated at the boundaries of segmentation targets, so we aimed to guide the model’s attention toward these critical boundary regions during training. To this end, we introduced the concept of contour-weighted learning, which enhanced the model’s discriminative ability at the boundaries.
This contour-weighted strategy was implemented through two complementary components: a contour-weighted cross-entropy loss that emphasizes learning of contour features, and a separable Dice loss that balances the contribution of contour and non-contour regions.
The proposed approach is model-independent and can be flexibly integrated into various segmentation frameworks. 
In addition, we introduced a lightweight yet effective network, PDANet, which integrated the partial decoder mechanism and receptive field blocks to enhance feature representation. 
Evaluations on three publicly available datasets with significant data imbalance demonstrated that our method not only outperformed several state-of-the-art approaches in terms of segmentation quality, but also consistently improved the accuracy of existing segmentation models when equipped with our new loss function.

Our main contributions are summarized as follows.
\begin{itemize}
\item[$\bullet$] We introduced the concept of contour-weighted for image segmentation and demonstrated its efficacy in addressing the problem of data imbalance.
\item[$\bullet$] We proposed a new compound loss function that integrates contour-weighted cross-entropy and separate Dice loss to mitigate volume disparities and sample distribution imbalances in segmentation tasks.
\item[$\bullet$] We designed a lightweight yet effective network to improve segmentation accuracy by combining partial decoder mechanisms and receptive field blocks, consistently achieving state-of-the-art performance compared to nine learning-based segmentation methods.
\item[$\bullet$] Our contour-weighted method is model-independent and can be flexibly integrated into arbitrary learning-based segmentation models, achieving average performance improvements of the nine existing popular models by 2.39\%, 1.44\%, and 3.88\% across three public datasets.
\end{itemize}

% This work extended our previous conference version \cite{huang2024contour} in both methods and experiments. We introduced PDANet to improve segmentation accuracy and explored the contour-weighted loss function theoretically and comprehensively. Experimentally, we engaged more public datasets for evaluation, and included more hyperparameter studies.

\section{Methods}
\label{sec: Methods}
To address the challenge of data imbalance in image segmentation tasks, we introduce the concept of contour-weighted maps that adaptively emphasize contour regions in the segmentation process. This approach enables deep networks to simultaneously capture global anatomical context while preserving fine-grained boundary details, effectively integrating both structural completeness and boundary precision in a unified framework. In addition, we design a lightweight network by introducing a partial decoder mechanism, which improves the segmentation accuracy while reducing the complexity of the model.

\subsection{Preliminary}
The section will present the main probability inequalities that underlie our approach.

\textbf{\textit{Theory:}} Our approach is based on the fact that: 
\textit{Let $f(x,y)=2xy/(x+y), f(x_1, y_1)=2x_1y_1/(x_1 + y_1), f(x_2, y_2)=2x_2y_2/(x_2 + y_2)$, if $x = x_1 + x_2, y=y_1 + y_2$, for any $x_1 \geq 0, x_2 \geq 0, y_1 \geq 0, y_2 \geq 0$, we have}
\begin{equation}
    f(x,y) \geq f(x_1, y_1) + f(x_2, y_2).
\end{equation}

The proof process is as follows:
\begin{assump}\label{assump}
    if $f(x,y) \textless f(x_1, y_1) + f(x_2, y_2)$ holds, we have
    \begin{equation}
        \frac{xy}{x+y} \textless \frac{x_1y_1}{x_1+y_1} + \frac{x_2y_2}{x_2+y_2}.
    \end{equation}
    Then we have
    \begin{equation}
        \begin{split}
        xy(x_1+y_1)(x_2)(y_2) &\textless x_1y_1(x+y)(x_2+y_2) \\
                            &+ x_2y_2(x+y)(x_1+y_1).
    \end{split}
    \end{equation}
    By further simplification, we have
    \begin{equation}
        \begin{split}
        xy(x_1y_2 + x_2y_1) &\textless x(x_1y_1y_2 + x_2y_1y_2) \\
                            &+ y(x_1x_2y_1 + x_1x_2y_2).
    \end{split}
    \end{equation}
    Because $x=x_1 + x_2, y=y_1 + y_2$, we can replace $x$ and $y$ in Eq.(4) to get
    % \begin{split}
    %     (x_1 + x_2)(y_1 + y_2)(x_1y_2 + x_2y_1) &\textless x_1x_1y_1y_2 + x_2x_1y_1y_2\\
    %     + x_1x_2y_1y_2 + x_2x_2y_1y_2 + (y_1 + y_2)(x_1x_2y_1 + x_1x_2y_2).
    % \end{split}
    % \end{equation}
    \begin{equation}
        \begin{split}
        x_1^2y_2^2 + x_2^2y_1^2 \textless 2x_1x_2y_1y_2.
    \end{split}
    \end{equation}
    Then, we have 
    \begin{equation}\label{eq: contradiction}
        (x_1y_2 - x_2y_1)^2 \textless 0.
    \end{equation}
\end{assump}
Eq.\ref{eq: contradiction} is a contradictory result, so the \textit{assumption} \ref{assump} is invalid.

Eventually, we have
\begin{equation}
    \frac{xy}{x+y} \geq \frac{x_1y_1}{x_1+y_1} + \frac{x_2y_2}{x_2+y_2}, 
\end{equation}
subjected to $x=x_1+x_2,y=y_1+y2, x_1\geq0, x_2\geq0, y_1\geq0,y_2\geq0$.

\subsection{Contour-Weighted Compound loss function}
We propose a novel compound loss function that combines contour-weighted cross-entropy loss and separable Dice loss to address the data imbalance problem caused by volume disparities and sample size disparities across different anatomical structures. The key idea is to leverage contour information to guide the network's attention towards boundary regions.

\subsubsection{Contour Extraction}
\label{sec: Contour Extraction}
\begin{figure}[t]
    \centering
    \includegraphics[width=\linewidth]{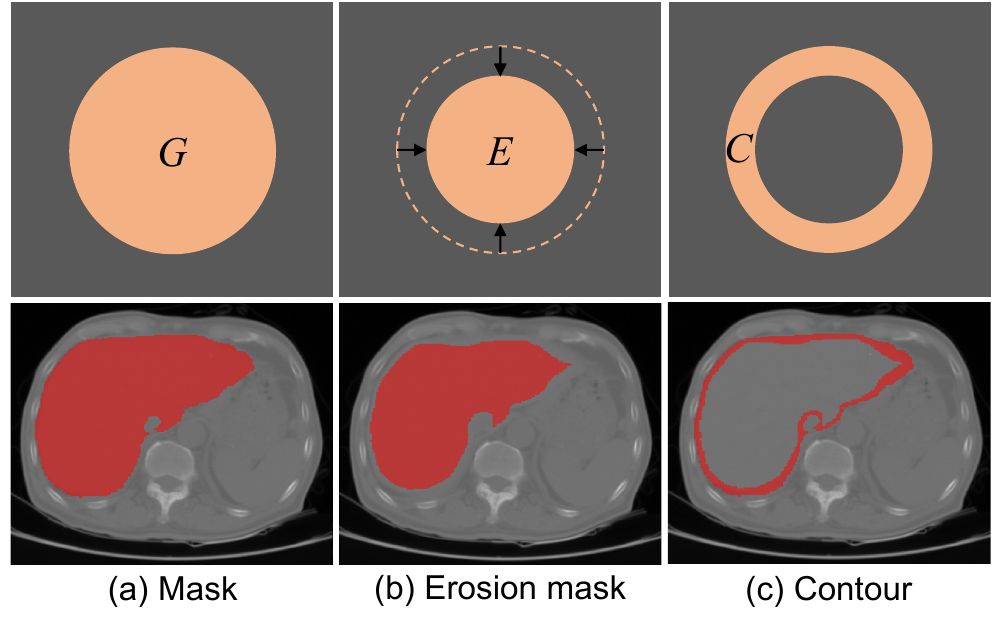}
    \caption{Illustration of our proposed contour-weighted map. We calculate the contours $C$ from the difference between the mask $G$ and its morphological erosion correspondence $E$ ($C = G-E$). The second row visualizes the contour calculation process.}
    \label{contour}
\end{figure}

We construct the contour-weighted map on the labeled image through morphological erosion operations, which can eliminate the boundary points of a connected region, thereby causing the boundary to shrink inward \cite{huang2024contour}. Fig. \ref{contour} illustrates the calculation strategy for the contour of the segmentation target. Specifically, we employ morphological erosion to cause the target boundary to shrink inward, resulting in a smaller target volume. Subsequently, the eroded target is subtracted from the original target to obtain the contour. This process is described as follows:
\begin{equation}
\label{equ: contour extraction}
    C = G - (G \ominus k), 
\end{equation}
where $G$ denotes the segmentation target, $\ominus$ represents the morphological erosion operation, and $k$ is the erosion kernel size. This process effectively captures the boundary regions that require special attention during training.

% FIG. 01
% =======

\subsubsection{Contour-Weighted Cross Entropy Loss}
\label{Contour Weighted Cross Entropy Loss}
Cross-entropy loss \cite{mao2023cross} is widely used in image segmentation. 
The binary cross-entropy loss is defined as
% We rewrite Eq.(\ref{bce loss}) as
\begin{equation}\label{ce loss}
    \mathcal{L}_{CE}^b = -\sum_{i=1}^{N}g_{i}\log p_{i}.
\end{equation}

For multi-category segmentation tasks, the cross-entropy loss is formulated as:
\begin{equation}\label{ce loss}
    \mathcal{L}_{CE}^m = - \frac {1}{M} \sum_{j=1}^{M} \sum_{i=1}^{N}g_{i}\log p_{i},
\end{equation}
where $M$ denotes the number of target categories to be segmented, and $N$ represents the total number of voxels in the image.

Building upon the extracted contour information, we design a contour-weighted cross-entropy (CWCE) loss that addresses the volume disparities. The standard cross-entropy loss treats all voxels equally, while our approach assigns higher weights to boundary regions, where accurate segmentation is most challenging and crucial. Our loss function is defined as:
\begin{equation}\label{ce loss}
    \mathcal{L}_{CE}^w = - \frac {1}{M} \sum_{j=1}^{M} \sum_{i=1}^{N} (\lambda{w_{c}}_{(i)} + 1)  g_{i}\log p_{i},
\end{equation}
where ${w_{c}}$ denotes the contour-weighted map we extracted using the Eq. (\ref{equ: contour extraction}). ${w_{c}}$ is a binary mask where 1 indicates the contour areas and 0 represents non-contour areas. $\lambda$ is a weighting factor.

\subsubsection{Separable Dice Loss}
\label{Separable Dice Loss}
The cross-entropy loss excels in voxel-level classification, while the Dice loss, derived from the Dice similarity coefficient, focuses on region-level consistency, thus providing complementary advantages. The standard Dice loss is formulated as:

\begin{equation}\label{dice loss}
    \mathcal{L}_{Dice} = 1 - 2 \frac{\sum_{i=1}^{N} p_{i} g_{i}} {\sum_{i=1}^{N} p_{i}^2+\sum_{i=1}^{N} g_{i}^2 + \epsilon },
\end{equation}
where $p_i$ and $g_i$ denote the $i$-th voxel from the predicted and ground truth segmentation, respectively, and $\epsilon$ is a small constant to avoid division by 0. 

To address the instability of Dice loss in multi-target segmentation, we propose a separable Dice loss (SDL) that decomposes the segmentation task into contour and non-contour components. This decomposition, illustrated in Fig. \ref{contour}, allows us to apply different weights to contour and non-contour regions. By calculating the consistency separately for these components, we achieve two key benefits: enhanced boundary accuracy through increased focus on contour regions, and improved data imbalance caused by volume variations among different targets.

\begin{equation}\label{loss contour}
    \mathcal{L}_{c} = 1- \frac {1}{M} \sum_{j=1}^{M} \frac{2 \sum_{i=1}^{N_1} {p_{i,j}^1}^2 {g_{i,j}^1}^2}{\sum_{i=1}^{N_1} {p_{i,j}^1}^2+\sum_{i=1}^{N_1} {g_{i,j}^1}^2 + \epsilon },
\end{equation}
\begin{equation}\label{loss noncontour}
    \mathcal{L}_{noc} = 1- \frac {1}{M} \sum_{j=1}^{M} \frac{2 \sum_{i=1}^{N_2} {p_{i,j}^2}^2 {g_{i,j}^2}^2}{\sum_{i=1}^{N_2} {p_{i,j}^2}^2+\sum_{i=1}^{N_2} {g_{i,j}^2}^2 + \epsilon },
\end{equation}

\begin{equation}\label{sDice loss}
    \mathcal{L}_{SDL} = \beta \mathcal{L}_{c} + (1-\beta)\mathcal{L}_{noc},
\end{equation}
where $M$ represents the total number of target regions, while $g^1$ and $g^2$ represent the ground truth masks for contour and non-contour regions, respectively, with $g^1 \cup g^2\in Mask$ and $N_1 + N_2 = N$, similarly, $p^1$ and $p^2$ denote the network's predictions for these regions. The parameter $\beta$ balances the contribution of contour and non-contour components in the final loss and takes a value from 0 to 1.

We combine the contour-weighted cross-entropy with the separable Dice loss (CWCD) to leverage the complementary strengths of both loss formulations. The final compound loss function is defined as:

\begin{equation}\label{ce loss}
    \mathcal{L} = \alpha\mathcal{L}_{SDL} + (1-\alpha)\mathcal{L}_{CE}^w
\end{equation}

% Building upon our preliminary exploration in \cite{huang2024contour}, this work extends the analysis and validation of our compound loss function across diverse medical image segmentation tasks, demonstrating its effectiveness in addressing data imbalance challenges.

% FIG. 
% =======
\begin{figure}[h]
    \centering
    \includegraphics[width=\linewidth]{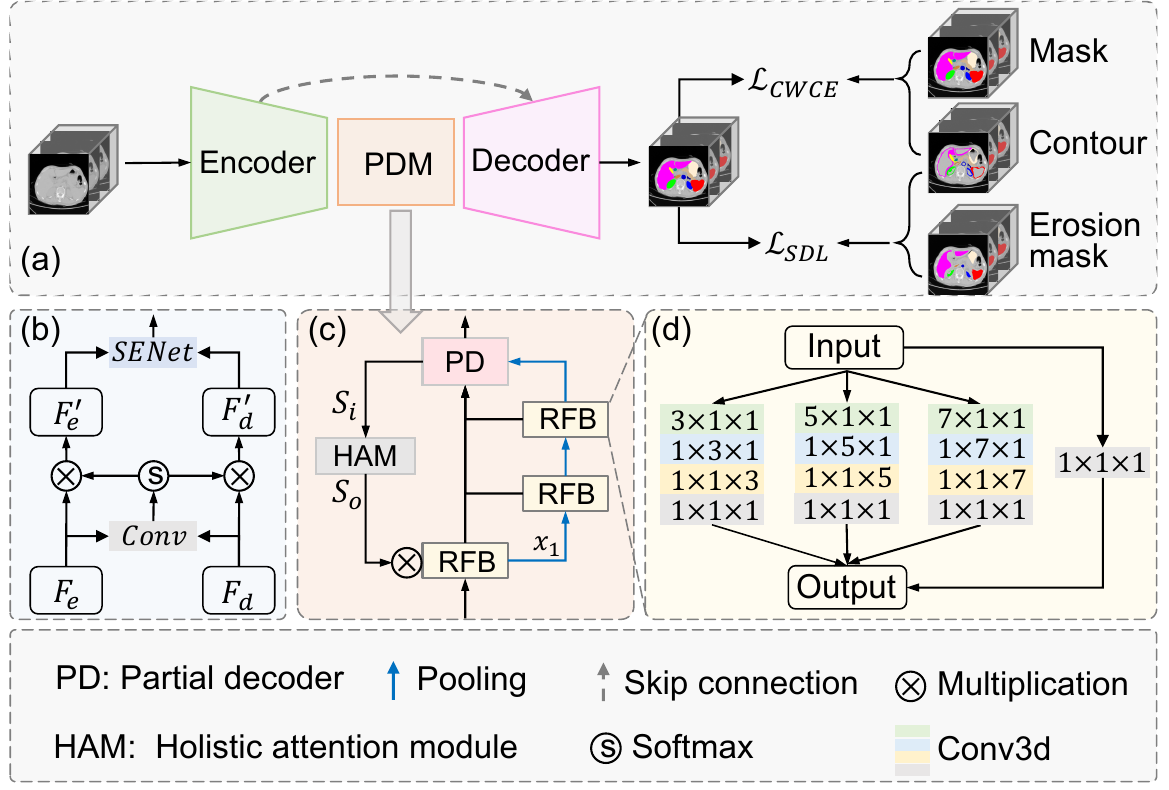}
     \caption{Overview of the proposed contour-weighted segmentation approach. (a) The main framework is an encoder-decoder structure with a partial decoder module (PDM) inserted in the middle. The loss function consists of two components: a contour-weighted cross-entropy loss ($\mathcal{L}_{SDL}$)  and a separable dice loss ($\mathcal{L}_{CWCE}$). Where both the segmented contour and the erosion mask are obtained through Eq. (\ref{equ: contour extraction}). (b) The channel-wise attention module (CWCA) is used to fuse the feature information from the encoder and decoder adaptively. (c) The partial decoder module (PDM) integrates multi-scale features through a U-Net-like structure and a holistic attention mechanism. (d) The receptive field block (RFB) employs different kernel sizes to capture multi-grained feature information.}
    \label{fig: network}
\end{figure}

\subsection{Network Architecture}
As illustrated in Fig. \ref{fig: network}(a), we propose PDANet, an efficient segmentation framework that integrates a lightweight encoder-decoder structure. The Squeeze-and-Excitation networks (SENet) \cite{hu2018squeeze} at each layer in the encoder adaptively select channel-wise feature responses. To enhance feature representation while maintaining computational simplicity, we introduce a partial decoder module (PDM) \cite{wu2019cascaded} that selectively processes deep features and generates attention-guided feature maps. The framework is further augmented by a channel-wise attention module (CWAM) in the decoder, which effectively bridges the encoder and decoder pathways, mitigating information loss typically associated with pooling operations.

\subsubsection{Backbone}
The backbone of PDANet follows a progressive feature learning strategy with three down/up sampling blocks. Starting with 16 kernels in the initial convolutional layer, we systematically double the feature channels during downsampling and halve them during upsampling, maintaining an efficient balance between feature representation capacity and computational cost. Each layer in the encoder contains a convolutional module and a SENet, and each layer in the decoder contains a CWCA module and a convolutional module.

\subsubsection{Partial Decoder Module}
The PDM serves as a key component in our architecture, designed to extract and process multi-grained feature information efficiently. As shown in Fig. \ref{fig: network}(c) and \ref{fig: network}(d), the PDM employs a cascade of Receptive Field Blocks (RFB) to generate rich multi-grained feature representations. These features are organized into three distinct resolution levels ($x_1$, $x_2$, and $x_3$) and processed through a U-Net-like partial decoder (PD) structure that effectively integrates information across scales. The output of the PD block is a saliency map.
To enhance boundary preservation and feature refinement, we incorporate a holistic attention module (HAM) \cite{wu2019cascaded} to expand the coverage area of the saliency map so that the initial segmentation is over-segmented as much as possible to avoid losing boundary information. The HAM operation is formulated as:
\begin{equation}
    S_o = MAX(f_{norm}(Conv_g(S_i,k)), S_i),
\end{equation}
where $Conv_g$ represents a convolutional layer with Gaussian kernel $k$ and zero bias, $f_{norm}(\cdot)$ normalizes the output to [0, 1], and $MAX(\cdot)$ enhances the saliency of prominent regions in $S_i$. We follow \cite{wu2019cascaded} to set the Gaussian kernel size to 11 with a standard deviation of 2.

The generated saliency map is then applied to refine the ($x_1$), followed by another round of feature extraction through RFBs and PD. The second output of PD serves as the final output of PDM. To optimize the feature representation, we enhance the RFB design with multi-scale kernels and introduce pooling layers between consecutive RFBs, progressively reducing spatial resolution.

\subsubsection{Channel-wise Attention Module}
The CWAM is designed to integrate the features from the encoder and decoder, thereby progressively refining the segmentation results. It comprises two key components: an adaptive channel fusion block \cite{huang2022isa} for dynamic feature weighting, and a SENet block for refined feature selection.
As shown in \ref{fig: network}(b), the CWAM processes encoder features ($F_e$) and decoder features ($F_d$) through a fusion pipeline. After channel-wise concatenation, the features undergo convolution and softmax operations to generate an attention weight tensor $W \in \mathcal{R}^{2 \times H \times W \times D}$. This tensor is then split into complementary weights $w_1$ and $w_2$, which modulate the importance of encoder and decoder features, respectively.
The weighted features are further refined through the SENet block to produce the final output feature maps $O \in \mathcal{R}^{C \times H \times W \times D}$. 
% This process can be formulated as follows:
% \begin{equation}
%     \begin{aligned}
%         W &= \mathcal{F}_{softmax}(\mathcal{F}_{conv}(F_e, F_d)), \\
%         W &= [w_1, w_2], \\
%         O &= \mathcal{F}_{senet}(w_1F_e, w_2F_d).
%     \end{aligned}
%     \label{eq: cwca}
% \end{equation}

\subsection{Datasets}
We assessed our proposed approach on three public datasets for volumetric segmentation, comprising 1) the abdominal multi-organ segmentation dataset (AMOS) \cite{ji2022amos}, brain tumor segmentation dataset (BraTS) \cite{baid2021rsna}, and pelvic bone fragments with injuries segmentation dataset (PENGWIN) \cite{liu2023pelvic}. All datasets involve multi-target segmentation and suffer from data imbalance.

\subsubsection{AMOS dataset}
The AMOS dataset has 221 CT scans containing 15 labels, including spleen (Sp), right kidney (RK), left kidney (LK), gallbladder (Ga), esophagus (Es), liver (Li), stomach (St), aorta (Ao), postcava (Po), pancreas (Pa), right adrenal gland (RAG), left adrenal gland (LAG), duodenum (Du), bladder (Bl) and prostate/uterus (P/U). 
All volumes were resampled into an isotropic voxel spacing of $1.0 \emph{mm} \times 1.0 \emph{mm} \times 1.0 \emph{mm}$ and were center-cropped to a size of $d\times 320\times 320$ voxels. $d$ is the number of slices along the axial direction. Then, we sliced each scan into a small volume of size $64\times 320\times 320$ with an overlap of 16. The 221 scans in the AMOS dataset were divided into 180, 20, and 21 scans for training, validation, and testing. 

\subsubsection{BraTS dataset}
The BraTS dataset contains 1251 multi-modal MRI scans (Flair, T1w, T2w, T1ce) with three labels: necrotic tumor core (NTC), peritumoral edema (ED), and enhancing tumor (ET). The training, validation, and testing sets contain 800, 200, and 251 scans, respectively. In this paper, we reported the segmentation results of the whole tumor (WT, the union of NTC, ED, and ET), enhancing tumor (ET), and tumor core (TC, the union of ET and NTC). All volumes were center-cropped to $128\times160\times160$ ($1.0 \emph{mm} \times 1.0 \emph{mm} \times 1.0 \emph{mm}$) voxels. 

\subsubsection{PENGWIN dataset}
The PENGWIN dataset contains 100 CT scans, and each bone anatomy (sacrum, left hipbone, right hipbone) has at most 6 fragments. All but one sample of a single bone anatomy had no more than four fragments. Therefore, we renamed the labels so that 1 to 4 are sacrum fragments, 5 to 10 are left hipbone fragments, and 11 to 14 are right hipbone fragments, resulting in a total of 14 segmentation categories.
All scans were resampled into an isotropic voxel spacing of $1.0 \emph{mm} \times 1.0 \emph{mm} \times 1.0 \emph{mm}$ and were center-cropped to a size of $d\times 300\times 400$ voxels. $d$ is the number of slices along the axial direction. Then, we cropped each volume into a small size $64\times 300\times 400$, with an overlap of 16. We use 70, 10, and 20 scans for training, validation, and testing. The fragment size within the PENGWIN dataset varies, and so does the number of samples per fragment. Labels 4, 9, and 10 each appear in only one sample. The corresponding samples were included in the training set, so these labels do not appear in the test set. 

% FIG. 
% =======
\begin{figure}[t]
    \centering
    \includegraphics[width=\linewidth]{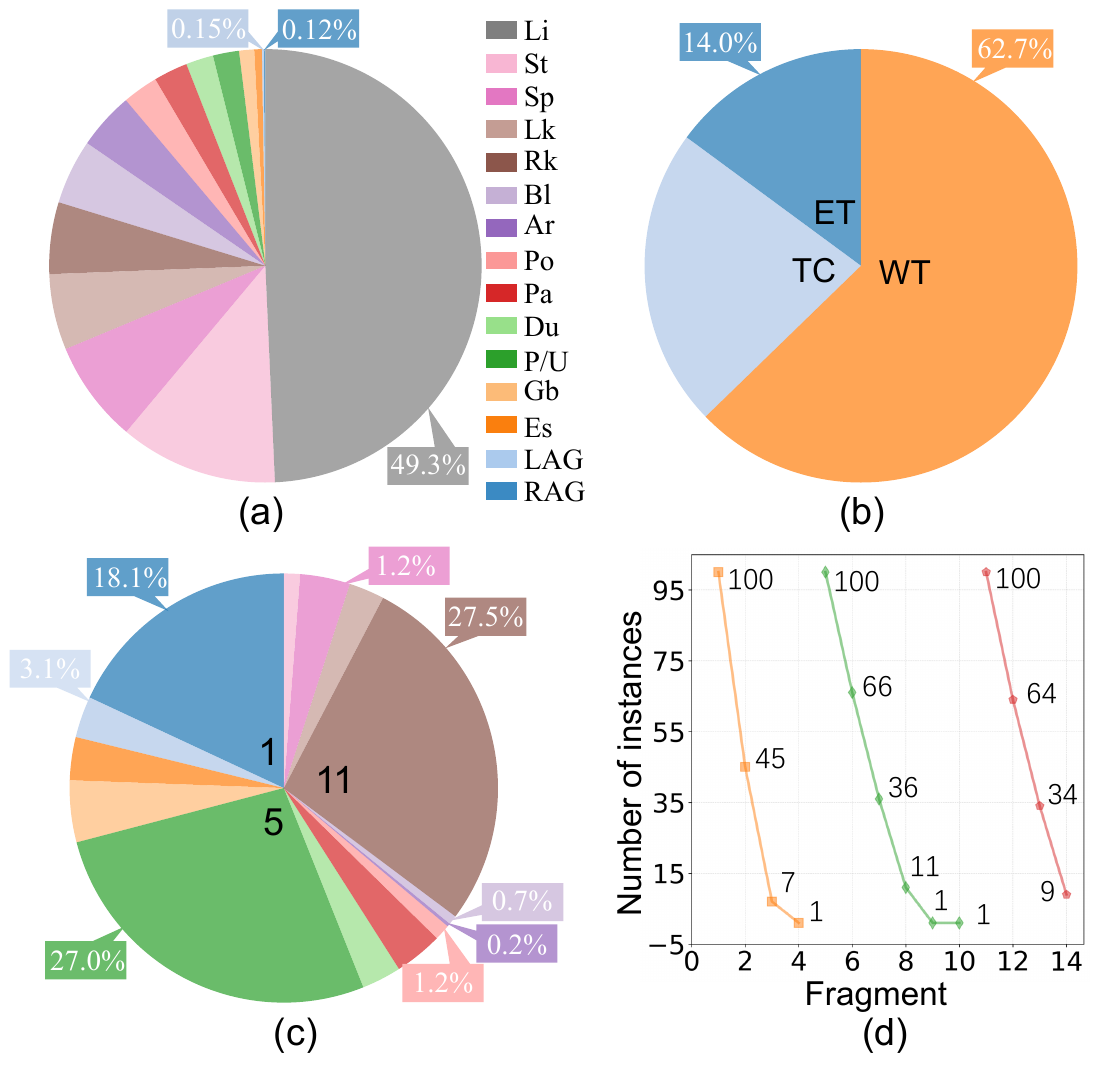}
    \caption{Data distribution bias. (a) The mean volume of different organs in the AMOS dataset. (b) The mean volume of different tissues in the BraTS dataset. (c) The mean volume of different fragments in the PENGWIN dataset. (d) The number of different segmentation fragments in the PENGWIN dataset. Orange lines indicate the sacrum, green indicates the left hipbone, and red indicates the right hipbone.}
    \label{fig: data_distribution}
\end{figure}

Fig. \ref{fig: data_distribution} depicts the distribution of the three datasets.
The AMOS dataset has 15 labels in total, and more than 90\% of the instances have 15 labels. A small number of instances have fewer than 15 labels, with a minimum of 13. The volumes of the 15 labels vary greatly, with an average difference of about 400 times between the average volume of the largest organ (Liver) and the smallest organ (right adrenal gland). 
Each instance in the BraTS dataset has 3 labels, but the volume of each label is different. The volume of the largest label (WT) is about 4 times that of the smallest one (ET). 
The PENGWIN dataset encompasses 14 fragments.
As can be seen in Fig. \ref{fig: data_distribution}(c), there is a significant imbalance in the volume of fragments in each bone anatomy. The difference in volume between the largest and smallest fragments in the sacrum is about 6-fold; about 135-fold in the left and 23-fold in the right hipbones. Fig. \ref{fig: data_distribution}(d) also shows the difference in the number of instances between each fragment.

\subsection{Implementation Details}
Our approach was implemented using PyTorch \cite{paszke2019pytorch} on an NVIDIA 4090 GPU. To fairly demonstrate the effectiveness of the proposed contour-weighted method, we train it separately on three datasets. All methods use the same data augmentation strategy. During training, the data enhancement methods are random rotation (angle range [0, 15] degrees with probability 0.2) and flips (horizontal or vertical flip with probability 0.2). The initial learning rate for all datasets is 3e-4, and the Adam \cite{kingma2014adam} optimizer is used. On the AMOS dataset, we carried out 50 epochs for training, and we halved the learning rate at the $20^{th}$ and $40^{th}$ epochs. On the BraTS and PENGWIN datasets, we trained the models for 100 epochs, with the learning rate linearly decaying to 1e-6 after 20 epochs.
The models with the best DSC performance on the validation set is selected for testing. The same contour extraction parameters are used for all three datasets, i.e., $k = 2, iter = 1$. The code can be accessed at \url{https://github.com/huangzyong/Contour-weighted-Loss-Seg}.

\begin{table}[t]
    \centering
    \caption{Influence of contour thickness on segmentation performance in terms of DSC (\%) based on 3D U-Net. Volume indicates the average size (voxels) of the organs. The size of the erosion kernel is set to 2, and \(iter\) denotes the number of iterations. Bold indicates the best results.}
    \label{tab: erosion parameter optimization}
    \resizebox{\linewidth}{!}{
    \begin{tabular}{cccccccc}
        \toprule
        \specialrule{0em}{0.2pt}{0.2pt}
        \hline
        Dataset & Organ &Volume & \textit{iter}=1 &\textit{iter}=2 & \textit{iter}=3 & \textit{iter}=4 & \textit{iter}=5 \\
        \hline
        \rule{0pt}{8pt} 
        	& Sp  &67634  &89.88	&90.64	&86.12	&89.08	&90.96  \\
                & RK  &47528  &91.08	&91.14	&91.12	&92.20	&91.77  \\  
                & LK  &50246  &91.90	&91.76	&91.55	&91.22	&91.74  \\
                & Ga  &10014  &60.55	&62.40	&58.17	&58.13	&57.89  \\ 
                & Es  &4995   &68.40	&67.51	&61.67	&58.59	&54.12  \\ 
                & Li  &437787 &94.40	&93.48	&94.29	&94.35	&94.30  \\
                & St  &104945 &80.96	&79.20	&80.35	&78.96	&78.23  \\
        AMOS    & Ao  &37865  &90.20	&91.15	&92.27	&90.74	&91.67  \\
                & Po  &22623  &84.11	&81.20	&83.48	&83.11	&83.50  \\
                & Pa  &23596  &67.35	&65.69	&63.37	&66.13	&61.75  \\
                & RAG &1059   &58.67	&58.25	&53.55	&50.04	&50.00  \\
                & LAG &1291   &53.47	&48.97	&45.46	&44.45	&36.44  \\
                & Du  &18240  &63.55	&65.05	&61.80	&60.58	&59.73  \\
                & Bl  &42971  &77.27	&70.22	&66.95	&67.31	&67.10  \\
                & P/U &17324  &63.24    &52.76	&54.29	&51.54	&62.11  \\
                \rowcolor[gray]{0.9} % 设为浅灰色
                \cellcolor[gray]{1}& Avg &- &\textbf{75.67} &73.96	&72.43	&71.76	&71.41  \\

        \hline
        \rule{0pt}{8pt} 
                &WT  &95967	 &86.00	&85.24	&85.64	&84.62	&83.14  \\
        BraTS   &TC	 &35753  &82.56	&82.52	&82.58	&82.58	&80.24  \\
                &ET	 &21447  &77.86	&77.81	&77.92	&77.18	&75.74  \\
                \rowcolor[gray]{0.9} % 设为浅灰色
                \cellcolor[gray]{1}&Avg &-  &\textbf{82.14}	&81.86	&82.05	&81.46	&79.71  \\     
        \hline
        \specialrule{0em}{0.2pt}{0.2pt}
        \bottomrule
    \end{tabular}}
\end{table}

\section{Results}
\label{sec: Results}

\subsection{Parameter Optimization} 
\subsubsection{Hyperparameter optimization for contour thickness}
We experimented with the AMOS and BraTS datasets to investigate the effect of contour thickness on segmentation performance. Specifically, we conducted experiments using the fixed erosion kernel size (\(k=2 \)) with different numbers of iterations, where larger iterations produce thicker contours. Table \ref{tab: erosion parameter optimization} shows that the highest average DSCs are achieved at \( iter=1 \) for both datasets (75.67\% for AMOS and 82.14\% for BraTS). As \( iter \) increases, the average DSC decreases, dropping to 71.41\% ($4.26\%\downarrow$) and 79.71\% ($2.43\%\downarrow$) respectively at $iter=5$.

\subsubsection{Hyperparameter optimization for contour-weighted compound loss function}
\begin{figure}[t]
    \centering
    \includegraphics[width=\linewidth]{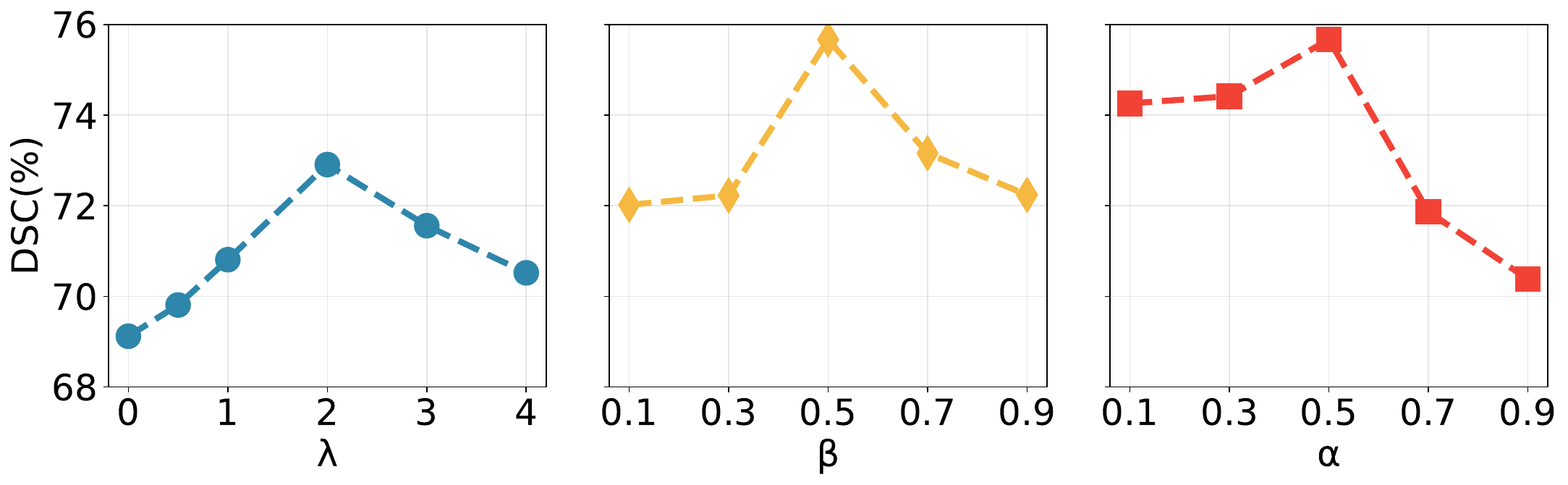}
    \caption{Influence of different parameters on the segmentation performance of 3D U-Net on the AMOS dataset. In the left column, $\alpha$ is set to 0. In the middle column, $\alpha$ is set to 0.5, and $\lambda$ is set to 2. In the right column, $\beta$ is set to 0.5, and $\lambda$ is set to 2.
}
    \label{fig: para alpaha beta lambda}
\end{figure}

Our CWCD loss function contains three adjustable parameters: $\alpha$, $\beta$, and $\lambda$. Fig. \ref{fig: para alpaha beta lambda} illustrates their effects on segmentation performance. 
Training with CWCE alone yielded relatively low DSC scores. When $\lambda$ is set to 0, CWCE degenerates into the standard cross-entropy loss, yielding the lowest DSC score (69.12\%). The integration of SDL significantly improved performance, with optimal results achieved at $\beta=0.5$, indicating equal importance of contour and non-contour components. Results also show that the weighting factor $\alpha$ for SDL should not be too large when using the compound loss function.
For subsequent experiments, we set $\alpha = 0.5$, $\beta = 0.5$, and $\lambda = 2$.

% FIG. 
% =======

\subsection{Ablation Study}
\subsubsection{Ablation study on contour-weighted compound loss function}
To evaluate each component's contribution, we conducted experiments using different loss combinations: cross-entropy (CE), contour-weighted cross-entropy (CWCE), dice loss (DL), and separable dice loss (SDL). 
Table \ref{tab: ablation CWCD} presents results on AMOS and BraTS datasets, revealing that: 1) both compound loss and contour-weighted strategy improve segmentation performance; 2) CWCE demonstrates more robust performance than SDL, particularly on the AMOS dataset; and 3) although DL and SDL show limited effectiveness on the AMOS dataset with numerous categories, they contribute positively when used in compound loss configurations.

On the AMOS dataset, Dice loss alone fails to segment most organs. In contrast, applying the proposed Separable Dice loss results in a marked improvement. Although a few organs remain unsegmented, the average DSC of the seven successfully predicted organs reaches 79.93\% (underlined). These results demonstrate that the contour-weighted strategy effectively improves segmentation performance under severe data imbalance.

\begin{table}[t]
    \centering
    \caption{Results of ablation study. The best performance per dataset in terms of DSC (\%) is highlighted in bold. A hyphen (-) indicates a complete failure in prediction.}
    \label{tab: ablation CWCD}
    \begin{tabular}{ccccccc}
        \toprule
        \specialrule{0em}{0.2pt}{0.2pt}
        \cline{1-7}
        Dataset & Organ & CE & CWCE & DL & SDL & CWCD\\ 
        \cline{1-7}
        	& Sp &86.79	&90.39	&-	&-	&89.88 \\
                & RK &91.93	&91.32	&-	&87.36	&91.08 \\
                & LK &88.52	&87.03	&-	&86.11	&91.90 \\
                & Ga &57.22	&58.59	&-	&-	&60.55 \\ 
                & Es &49.76	&62.83	&-	&-	&68.40 \\ 
                & Li &93.50	&90.52	&9.36	&92.38	&94.40 \\
                & St &81.94	&76.85	&-	&76.32	&80.96  \\
        AMOS    & Ao &90.35	&90.63	&-	&87.67	&90.20  \\
                & Po &82.63	&79.55	&-	&-	&84.11  \\
                & Pa &64.91	&64.95	&-	&-	&67.35  \\
                & RAG&50.83	&57.59	&53.76	&-	&58.67  \\
                & LAG&35.66	&59.66	&-	&-	&53.47  \\
                & Du &60.21	&66.39	&-	&-	&63.55  \\
                & Bl &61.61	&60.15	&-	&70.62	&77.27  \\
                & P/U &40.97	&52.71	&-	&59.06	&63.24  \\
                \rowcolor[gray]{0.9} % 设为浅灰色
                \cellcolor[gray]{1}& Avg &69.12	&72.91	&31.56	&\underline{79.93}	&\textbf{75.67} \\

        \cline{1-7}
        \rule{0pt}{8pt} 
                &WT	&81.87	&84.15	&76.15	&78.30	&86.00  \\
        BraTS   &TC	&76.33	&79.29	&64.17	&68.14	&82.56  \\
                &ET	&73.27	&75.08	&72.59	&76.47	&77.86  \\
                \rowcolor[gray]{0.9} % 设为浅灰色
                \cellcolor[gray]{1}&Avg&77.16	&79.51	&70.97	&74.31	&\textbf{82.14}  \\     
        \cline{1-7}
        \specialrule{0em}{0.2pt}{0.2pt}
        \bottomrule
    \end{tabular}
\end{table}

% FIG. 
% =======
\begin{figure}[t]
    \centering
    \includegraphics[width=\linewidth]{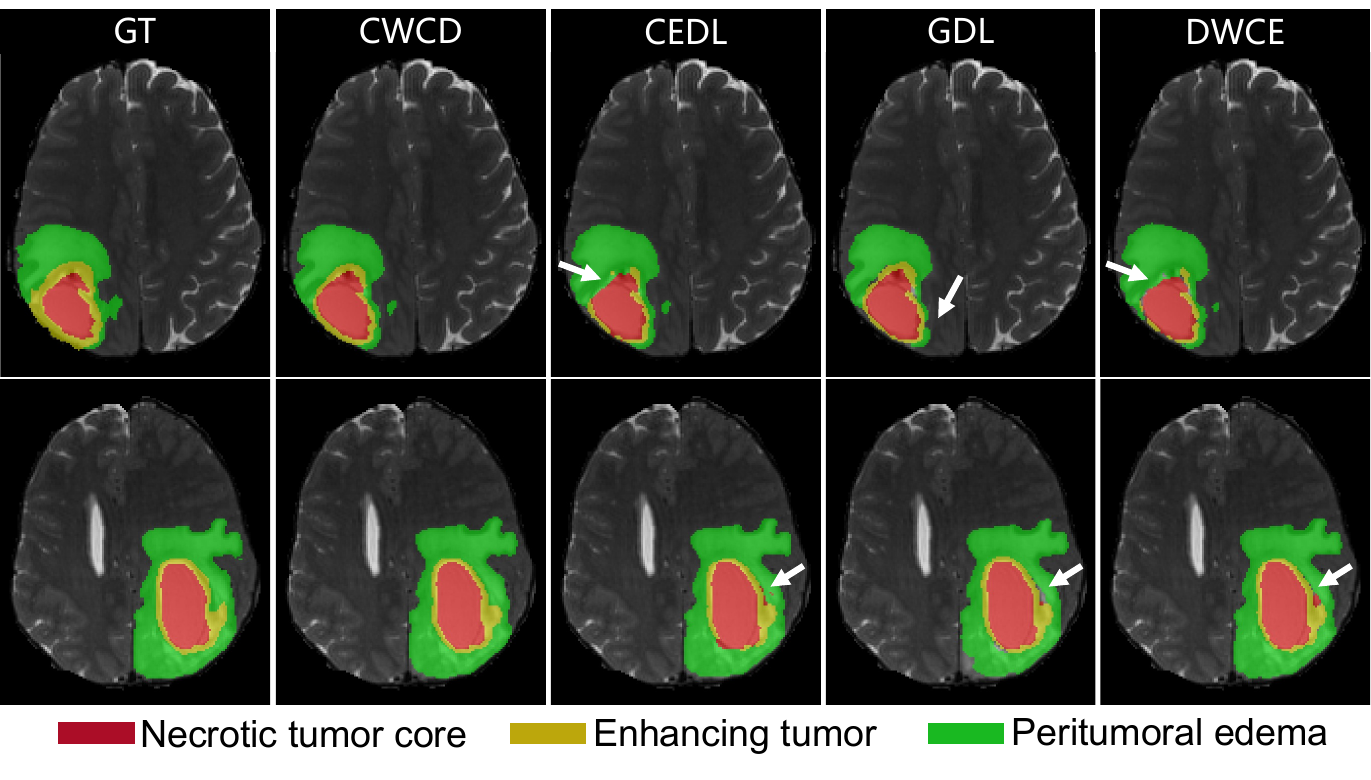}
    \caption{Qualitative comparison of different loss functions based on 3D U-Net on the BraTS dataset. The whole tumor (WT) encompasses a union of red, yellow, and green regions. The tumor core (TC) includes the union of red and yellow regions. The enhancing tumor (ET) denotes the yellow region.}
    \label{fig: brats visual}
\end{figure}

% FIG. 
% =======
\begin{figure}[h]
    \centering
    \includegraphics[width=\linewidth]{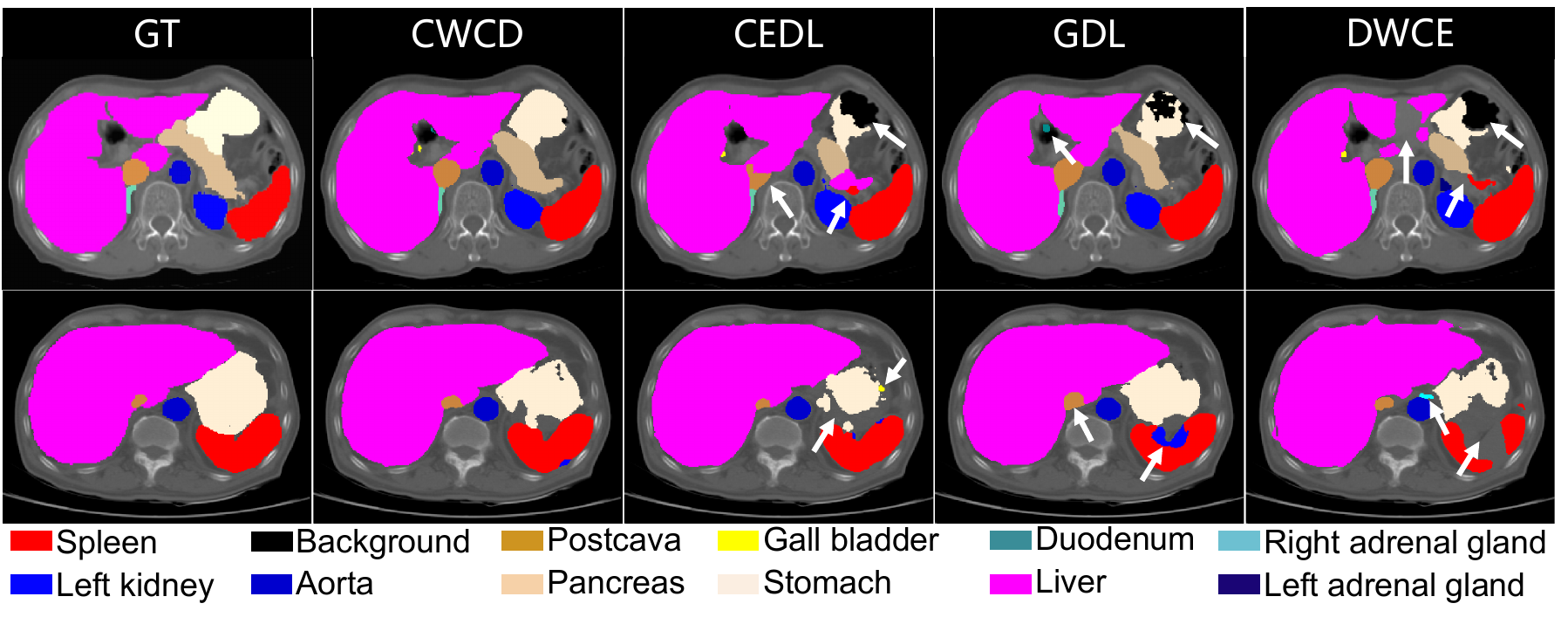}
    \caption{Qualitative comparison of different loss functions based on 3D U-Net on the AMOS dataset.}
    \label{fig: abdominal visual}
\end{figure}

%%%
Based on the contour extraction strategy outlined in Section~\ref{sec: Contour Extraction}, along with the results presented in Table \ref{tab: erosion parameter optimization}, Table \ref{tab: ablation CWCD}, and Fig. \ref{fig: para alpaha beta lambda}, we derived the following observations:
1) When the volume of the segmented organ or tissue is very small, a higher value of $iter$ causes the extracted contour to occupy a disproportionately larger proportion of the segmented target. This causes segmentation errors to concentrate in the contour region, while the non-contour region's contribution to the loss becomes negligible. Consequently, $\mathcal{L}_c$ approximates $\mathcal{L}_{Dice}$, and $\mathcal{L}_{noc}$ loses its effectiveness.
2) Conversely, when the volume of the segmented target is large, the extracted contour occupies only a small proportion of the entire segmented target. The large segmentation errors persist in the contour region, making it difficult to effectively minimize $\mathcal{L}_c$. 
In both scenarios, SDL tends to degenerate into the standard Dice loss, thereby undermining its advantage of addressing the data imbalance. As demonstrated in Eq. (\ref{sDice loss}) and Fig. \ref{fig: para alpaha beta lambda}, both $\mathcal{L}_c$ and $\mathcal{L}_{noc}$ are important for improving segmentation performance.
Therefore, we need to make a trade-off among multiple segmentation targets and use appropriate parameters \(iter\) to achieve the best results.

\subsubsection{Ablation Study on Network}
We evaluated the contributions of SENet, PDM, and CWAM through ablation experiments on PDANet. As shown in Table \ref{tab: ablation network}, the baseline network alone achieved a DSC of 79.78\%. The integration of SENet and PDM led to a performance increase of 3.03\%, while the subsequent addition of CWAM significantly improved to 84.50\% (a total gain of +4.72\%). These results demonstrate the effectiveness of each architectural component in improving segmentation performance.

\begin{table}[t]
  \caption{Results of the ablation study for the SENet, PDM, and CWAM modules. The results are expressed as the mean DSC (\%) score on the AMOS dataset.}
  \label{tab: ablation network}
  \centering
  % \resizebox{\linewidth}{!}{
        \begin{tabular}{c|cccc}
        \hline 
        \specialrule{0em}{0.2pt}{0.2pt}
        \hline 
        Methods  & SENet & PDM & CWAM & DSC (\%) $\uparrow$   \\ 
        \hline 
        Model 1  &\ding{55}    &\ding{55}    &\ding{55}      &79.78  \\
        Model 2  &$\checkmark$ &\ding{55}    &\ding{55}      &80.32  \\
        Model 3  &$\checkmark$ &$\checkmark$ &\ding{55}      &82.81  \\
        Model 4  &$\checkmark$ &$\checkmark$ &$\checkmark$   &\textbf{84.50}  \\
        \hline 
        \specialrule{0em}{0.2pt}{0.2pt}
        \hline
      \end{tabular}
       % }
\end{table}

\subsection{Comparison with Different Loss Functions}
To assess the performance of our proposed compound loss function, we compared it with three related loss functions: Generalized Dice Loss (GDL) \cite{sudre2017generalised}, Distance-Weighted Cross Entropy (DWCE) \cite{caliva2019distance}, and Combo Loss (CEDL) \cite{taghanaki2019combo}. 

We trained the 3D U-Net using different loss functions on the AMOS and BraTS datasets with the same protocol. The results are shown in Table \ref{tab: different loss}. The proposed loss function achieved superior performance on both datasets in terms of DSC. Our loss function outperforms the second-best (CEDL) by 4.64\% on the AMOS dataset. Furthermore, our CWCD loss outperformed GDL (which ranks second) on the BraTS dataset (82.14\% vs 80.70\%). These qualitative results suggest that our proposed loss function is more effective in addressing data imbalance.

Fig. \ref{fig: brats visual} and Fig. \ref{fig: abdominal visual} present the qualitative evaluations of two representative slices in the AMOS and BraTS datasets. These visual slices demonstrate that our CWCD loss produced the most accurate segmentation results compared to its three competing counterparts.

\begin{table}[t]
    \centering
    \caption{Segmentation results based on 3D U-Net using different loss functions on the AMOS and BraTS datasets, respectively, in terms of DSC ($\%$). The best results are highlighted in bold. *: \textit{p} $\textless$ 0.05, in comparison to our CWCD.}
    \label{tab: different loss}
    \begin{tabular}{cccccc}
        \toprule
        \specialrule{0em}{0.2pt}{0.2pt}
        \cline{1-6}
        \specialrule{0em}{0.2pt}{0.2pt}
        % Dataset & Organ & GDL\cite{sudre2017generalised} & DWCE\cite{ronneberger2015u} & CEDL\cite{taghanaki2019combo} & CWCD \\ 
        Dataset & Organ & GDL & DWCE & CEDL & CWCD \\ 
        % Dataset & Organ & GDL & DWCE & CEDL & CWCD \\ 
        \cline{1-6}
        	& Sp  &89.65	&83.51	 &88.31   &89.88  \\
                & RK  &92.41	&86.61   &89.61	  &91.08  \\
                & LK  &91.12	&88.47   &88.75   &91.90  \\
                & Ga  &55.27	&61.53	 &64.16   &60.55  \\ 
                & Es  &47.94	&66.34	 &51.70	  &68.40  \\ 
                & Li  &94.79	&89.95	 &94.20	  &94.40  \\
                & St  &74.51	&68.21	 &77.15	  &80.96  \\
        AMOS    & Ao  &90.23	&89.26	 &90.51	  &90.20  \\
                & Po  &83.06	&80.34	 &82.88   &84.11  \\
                & Pa  &62.46	&61.52	 &59.74	  &67.35  \\
                & RAG &53.95	&60.87	 &43.97	  &58.67  \\
                & LAG &36.19	&62.08	 &35.49   &53.47  \\
                & Du  &61.56	&59.11   &59.42	  &63.55  \\
                & Bl  &68.79	&63.81   &71.20	  &77.27  \\
                & P/U &55.01	&26.71   &68.42   &63.24  \\
                \rowcolor[gray]{0.9} % 设为浅灰色
                \cellcolor[gray]{1}& Avg &70.46*	&69.89*   &71.03*	  &\textbf{75.67} \\

        \cline{1-6}
        \rule{0pt}{8pt} 
                &WT	 &80.64	 &82.56	 &84.11	 &86.00  \\
        BraTS   &TC	 &82.92	 &80.81  &78.68	 &82.56  \\
                &ET	 &78.55	 &75.72  &74.47	 &77.86  \\
                \rowcolor[gray]{0.9} % 设为浅灰色
                \cellcolor[gray]{1}&Avg &80.70*	 &79.70*	 &79.09*	 &\textbf{82.14}  \\     
        \cline{1-6}
        \specialrule{0em}{0.2pt}{0.2pt}
        \bottomrule
    \end{tabular}
\end{table}

\begin{table*}[h]
    \centering
    \caption{Comparison of different segmentation models trained with the original loss function and our CWCD loss on all datasets (original loss / CWCD). Our PDANet was trained only once using the CWCD loss. The better performance relative to the original loss is indicated in bold. Red indicates the best result of all the methods, and underline indicates the second-best. *: \textit{p} $\textless$ 0.05, comparison between original loss and our CWCD loss.}
    \label{tab: different models}
    \resizebox{\linewidth}{!}{
    \begin{tabular}{c|lll|lll|lll|l}
        \toprule
        \specialrule{0em}{0.2pt}{0.2pt}
        \cline{1-11}
        \specialrule{0em}{0.2pt}{0.2pt}
        \multirow{2}{1cm}{Methods} & \multicolumn{3}{c|}{AMOS} & \multicolumn{3}{c|}{BraTS} & \multicolumn{3}{c|}{PENGWIN} &\multirow{2}{1cm}{$\#$Params} \\
        \cline{2-10}
        % \cmidrule(lr){2-4} \cmidrule(lr){5-7} \cmidrule(lr){8-10} 
        \specialrule{0em}{0.2pt}{0.2pt}
        &\multicolumn{1}{c}{DSC (\%) $\uparrow$} & \multicolumn{1}{c}{HD95 $\downarrow$} & \multicolumn{1}{c|}{ASSD $\downarrow$} &\multicolumn{1}{c}{DSC (\%) $\uparrow$} & \multicolumn{1}{c}{HD95 $\downarrow$} & \multicolumn{1}{c|}{ASSD $\downarrow$} &\multicolumn{1}{c}{DSC (\%) $\uparrow$} & \multicolumn{1}{c}{HD95 $\downarrow$} & \multicolumn{1}{c|}{ASSD $\downarrow$} \\ 
        \hline
        \rule{0pt}{8pt} 

        3D U-Net \cite{ronneberger2015u} & 71.03 / \textbf{75.67}*	&7.23 / \textbf{6.63}*	&2.00 / \textbf{1.94}	& 79.09 / \textbf{82.14}*	&7.64 / \textbf{6.86}* &1.63 / \textbf{\textcolor{red}{1.48}}* &68.63 / \textbf{73.60}* &40.04 / \textbf{33.56}* &11.32 / \textbf{8.28}* &1.87M\\

        DeepLabV3 \cite{chen2017rethinking} & 78.61 / \textbf{81.32}*	&4.71 / \textbf{4.35}* &1.31 / \textbf{\textcolor{red}{1.26}}	& 81.30 / \textbf{82.24}*	&8.22 / \textbf{7.16}*	&2.26 / \textbf{1.83}* &74.29 / \textbf{76.41}* & 27.46 / \textbf{23.82}* & 8.03 / \textbf{7.09}* &6.25M\\

        3D UX-Net \cite{lee20223d} & 81.52 / \textbf{83.62}*	&5.02 / 4.92 &1.75 / \textbf{1.71}	& 86.99 / \textbf{\underline{87.32}}	&6.36 / \textbf{5.79}*	&2.06 / \textbf{1.84}* &73.63 / \textbf{77.27}* &29.74 / \textbf{23.37}* &7.38 / \textbf{6.34}* &53.00M\\
    
        RepUX-Net \cite{lee2023scaling} & 83.30 / \textbf{\underline{84.02}}*	&4.60 / \textbf{4.49}	&1.71 / \textbf{1.46}*	& 85.31 / \textbf{86.52}*	&7.85 / \textbf{6.74}*	&2.59 / \textbf{2.20}*  &77.72 / \textbf{\underline{78.95}}* & 22.33 / 23.07 &6.42 / \textbf{6.11} &65.80M\\

        UNETR \cite{hatamizadeh2022unetr} & 61.16 / \textbf{63.72}*	&15.66 / \textbf{14.12}* 	&4.87 / \textbf{3.84}* 	& 72.72 / \textbf{75.62}*	&12.20 / \textbf{10.77}*	&3.87 / \textbf{3.19}* &55.71 / \textbf{64.66}* &47.83 / \textbf{36.73}* &12.12 / \textbf{9.36}* &104.05M\\
        
        Slim-UNETR \cite{pang2023slim} & 67.75 / \textbf{68.91}*	&10.87 / \textbf{10.85}	&3.31 / 3.39	& 81.50 / \textbf{82.01}	&7.23 / \textbf{6.62}*	&2.00 / \textbf{1.94} & 70.02 / \textbf{74.79}* & 36.63 / \textbf{28.48}* & 8.87 / \textbf{8.09}* &1.79M \\

        SwinUNETR \cite{hatamizadeh2021swin} & 82.94 / \textbf{83.89}*	&5.26 / \textbf{4.92}*	&1.79 / \textbf{1.72}*	& 85.65 / \textbf{86.88}*	&6.18 / \textbf{\textcolor{red}{5.62}}*	&2.03 / \textbf{\underline{1.77}}* &74.04 / \textbf{77.29}* &26.63 / \textbf{21.85}* &7.11 / \textbf{\underline{5.83}}* &61.99M\\

        nnFormer \cite{zhou2023nnformer} & 76.37 / \textbf{81.77}*	&4.98 / \textbf{\textcolor{red}{4.02}}*	&1.61 / \textbf{1.52}*	& 84.59 / \textbf{85.35}*	&7.00 / \textbf{6.77}*	&2.34 / 2.37  &73.31 / \textbf{75.82}* & 23.91 / \textbf{20.47}* & 6.39 / \textbf{5.92} &149.17M\\

        VSmTrans \cite{liu2024vsmtrans} &83.07 / \textbf{83.60}* &5.26 / \textbf{5.23} & 1.81 / \textbf{1.75} &84.80 / \textbf{86.82}*  &7.53 / \textbf{6.84}* & 2.50 / \textbf{2.29}* &75.53 / \textbf{77.25}* & 24.78 / \textbf{\underline{20.03}}* & 7.31 / \textbf{6.04}* &49.86M\\
        \rowcolor[gray]{0.9} % 设为浅灰色
        Ours &\multicolumn{1}{c}{\textcolor{red}{84.50}}	&\multicolumn{1}{c}{\underline{4.08}}	&\multicolumn{1}{c|}{\underline{1.51}}	&\multicolumn{1}{c}{\textcolor{red}{88.83}} &\multicolumn{1}{c}{\underline{5.75}} &\multicolumn{1}{c|}{1.81} &\multicolumn{1}{c}{\textcolor{red}{81.86}} &\multicolumn{1}{c}{\textcolor{red}{17.37}} &\multicolumn{1}{c|}{\textcolor{red}{4.53}} &31.48M\\
\hline
\specialrule{0em}{0.2pt}{0.2pt}
\bottomrule
\end{tabular}}
\end{table*}

Table~\ref{tab: erosion parameter optimization} and Table \ref{tab: different loss} reveal the robust performance of CWCD loss across varying contour extraction parameters. Notably, even with a challenging setting of $iter = 5$, CWCD maintained strong performance, achieving a DSC of 71.41\% on AMOS (surpassing CEDL) and 79.71\% on BraTS (outperforming both DWCE and CEDL). This demonstrates the stability and adaptability of our method under different parameter configurations.

Table \ref{tab: different loss} reveals an interesting pattern in segmentation performance across varying organ volumes: while DWCE excelled in segmenting small-volume structures (LGA, RGA, ES), its performance declined when dealing with large-volume organs (Li, Sp, St), lagging behind other methods.
In contrast, CWCD maintained consistent performance across both small and large structures, highlighting its effectiveness in addressing the volume-based data imbalance challenge.

\subsection{Comparison with Different State-of-the-Art Methods}
To comprehensively evaluate our contour-weighted method, we conducted extensive comparative experiments with nine state-of-the-art segmentation architectures (4 CNN-based and 5 Transformer-based). These models represent diverse architectural paradigms and complexity levels, ranging from lightweight networks to computationally intensive frameworks: 3D U-Net \cite{ronneberger2015u}, DeepLabV3 \cite{chen2017rethinking}, 3D UX-Net \cite{lee20223d}, RepUX-Net \cite{lee2023scaling}, UNETR \cite{hatamizadeh2022unetr}, SwinUNETR \cite{hatamizadeh2021swin}, Slim-UNETR \cite{pang2023slim}, nnFormer \cite{zhou2023nnformer}, and VSmTrans \cite{liu2024vsmtrans}. 
Each architecture underwent two independent training processes: one with its original loss function and another with our CWCD loss. The model's hyper-parameters are kept constant for both training, except for the loss function. The original loss functions varied across architectures: 3D U-Net employed weighted cross-entropy loss, DeepLabV3 utilized standard cross-entropy, Swin-UNETR implemented soft Dice loss, and Slim-UNETR combined soft Dice loss with focal loss. The remaining models used a hybrid approach combining soft Dice loss and cross-entropy loss. Our proposed PDANet, trained exclusively with the CWCD loss.

% FIG. 
% =======
\begin{figure}[t]
    \centering
    \includegraphics[width=\linewidth]{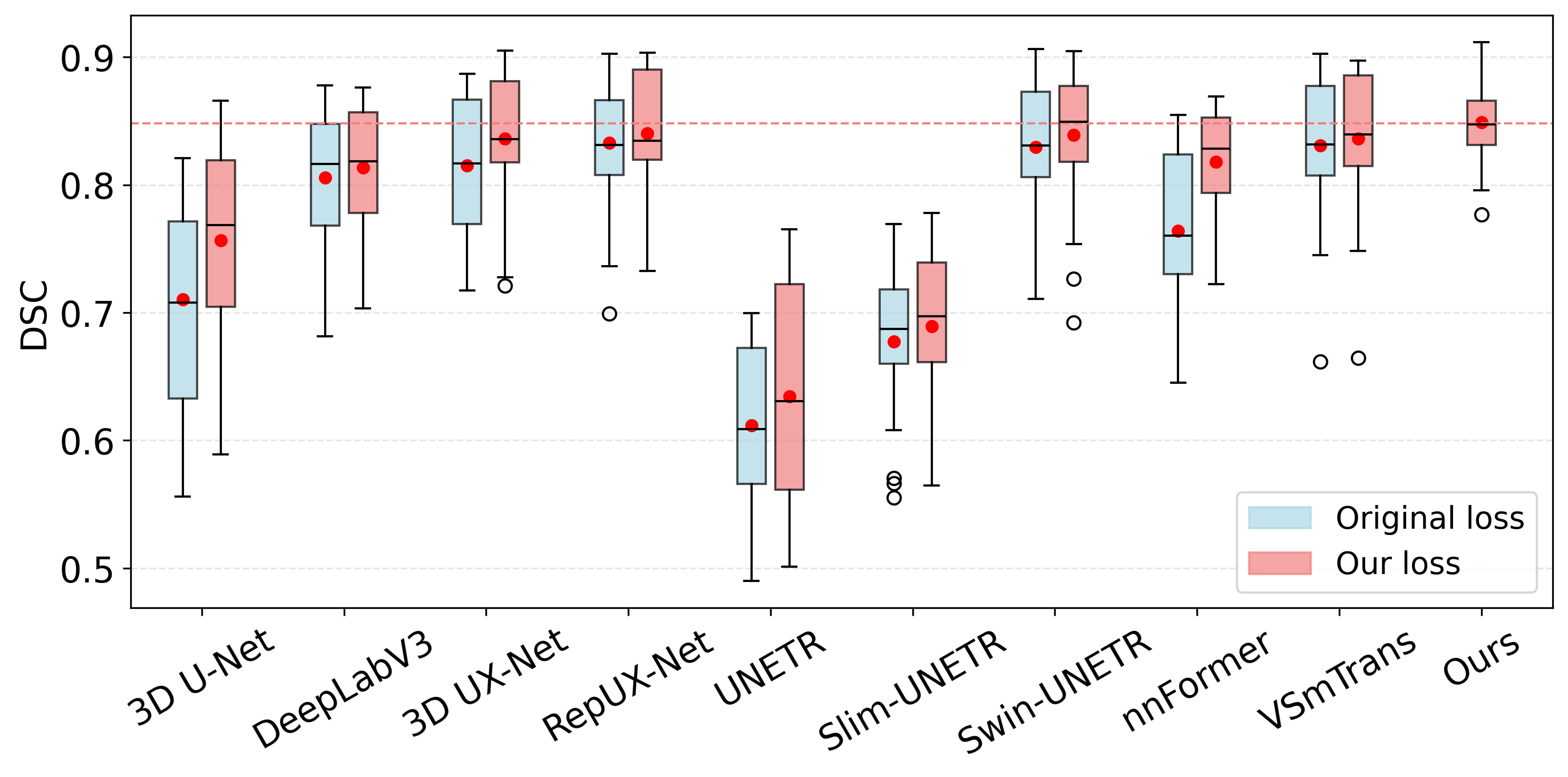}
    \caption{Quantitative comparison of different models using original loss and our loss on the AMOS dataset. Our PDANet was trained only once using CWCD loss. Solid red dots indicate the mean value.}
    \label{fig: box-amos-dsc}
\end{figure}

% FIG. 
% =======
\begin{figure}[t]
    \centering
    \includegraphics[width=\linewidth]{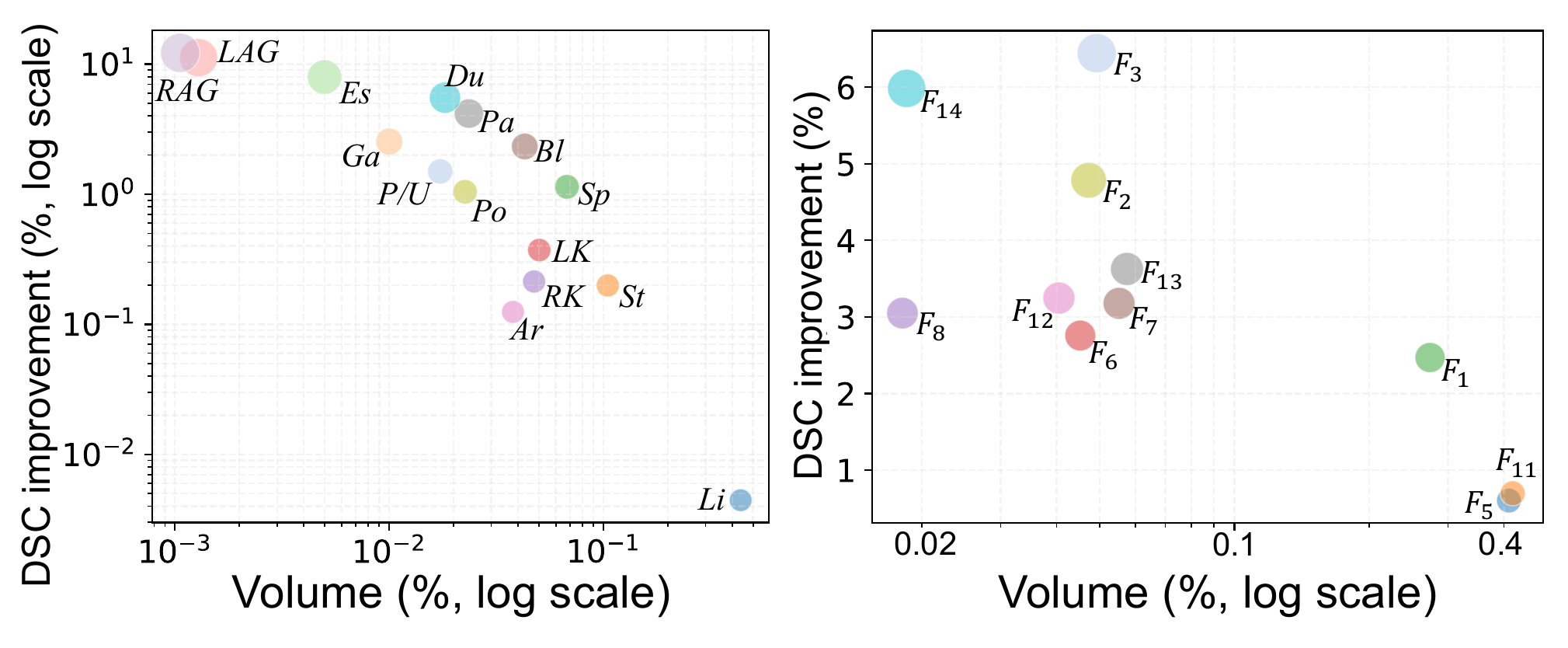}
    \caption{Segmentation performance (DSC improvement) versus volume fraction for different anatomical structures using the proposed CWCD loss on the AMOS (left) and PENGWIN (right) datasets. The values on the y-axis represent the mean DSC improvement of the nine comparison methods on the organs/fragments.}
    \label{fig: gain-amos-pengwin-dsc}
\end{figure}

\subsubsection{Assessment of accuracy}
Table \ref{tab: different models} demonstrates our approach achieved superior DSC scores of 84.50\%, 88.83\%, and 81.86\% on AMOS, BraTS, and PENGWIN datasets, respectively. Table~\ref{tab: fragment segmentation} shows the segmentation results for different fragments on the PENGWIN dataset, and our method achieved consistently optimal results on all fragments. Notably, on the PENGWIN dataset, our method exhibited exceptional performance across all three evaluation metrics (DSC, HD95, and ASSD), significantly outperforming comparison methods. Statistical analysis using the Wilcoxon signed-rank test revealed significant improvements ($p<0.05$) in segmentation performance when models were trained with CWCD loss compared to their original loss functions. 

Fig. \ref{fig: box-amos-dsc} presents a comparative analysis of model performance distributions, where the red dashed line representing PDANet's mean performance consistently exceeds all baseline methods. The box plots reveal that our method not only improved mean performance but also reduced standard deviation, indicating enhanced stability.

\begin{table*}[t]
  \caption{Segmentation results across different fragments on the PENGWIN dataset, reported as the mean DSC (\%). The better performance per comparison method is indicated in bold. Red indicates the best result of all the methods, and underline indicates the second-best. (Original loss / Our CWCD loss)}
  \label{tab: fragment segmentation}
  \centering
  \resizebox{\linewidth}{!}{
        \begin{tabular}{c|*{11}{c}}
        \hline 
        \specialrule{0em}{0.2pt}{0.2pt}
        \hline 
        Methods  & fragment-1 & fragment-2 & fragment-3 & fragment-5 & fragment-6 & fragment-7 & fragment-8 & fragment-11 & fragment-12 & fragment-13 & fragment-14   \\ 
        \hline 
        Volume & 174676 & 30104 &31343 &260912 &28804 &35067 &11550 &266266 & 25870 &36467 &11783 \\
        3D U-Net \cite{ronneberger2015u}  &85.22 / \textbf{85.37} &56.82 / \textbf{59.47} &49.21 / \textbf{56.32} &88.06 / 87.52 &54.63 / \textbf{58.79} &44.02 / 43.90 &50.31 / \textbf{52.93} &93.06 / 92.17 &55.68 / \textbf{59.28} &43.37 / \textbf{48.89} &45.02 / \textbf{51.21}  \\
        DeepLabV3 \cite{chen2017rethinking} &86.88 / \textbf{91.01} &60.01 / \textbf{63.07} &55.39 / \textbf{\underline{57.97}} &88.85 / \textbf{89.28} &57.92 / \textbf{59.05} &48.92 / \textbf{51.71} &53.38 / 52.27 &90.58 / \textbf{\underline{94.26}} &57.27 / \textbf{61.60} &53.02 / \textbf{55.89} &49.46 / 47.17\\
        3D UX-Net \cite{lee20223d}  &86.45 / \textbf{90.10} &62.03 / \textbf{\underline{66.04}} &51.17 / \textbf{55.79} &88.88 / \textbf{90.36} &59.30 / \textbf{61.52} &54.02 / \textbf{\underline{55.46}} &49.08 / \textbf{56.74} &92.46 / \textbf{93.64} &64.02 / 63.23 &51.09 / \textbf{\underline{54.57}} &50.07 / \textbf{\underline{56.67}} \\
        RepUX-Net \cite{lee2023scaling} &89.37 / \textbf{\underline{91.97}} &63.36 / \textbf{64.82} &58.53 / 57.89 &90.48 / \textbf{\underline{91.10}} &63.84 / \textbf{\underline{65.17}} &52.40 / \textbf{54.78} &57.28 / \textbf{\underline{60.12}} &93.90 / 93.88 &66.93 / \textbf{\underline{68.37}} &53.70 / 52.63 &48.95 / \textbf{52.18}\\
        UNETR \cite{hatamizadeh2022unetr} &82.74 / \textbf{85.97} &30.28 / \textbf{41.39} &17.28 / \textbf{45.01} &85.10 / \textbf{86.99} &35.01 / \textbf{39.43} &28.91 / \textbf{34.47} &18.29 / \textbf{25.68} &89.50 / \textbf{90.98} &20.39 / \textbf{38.21} &30.92 / \textbf{35.62} &0.00 / \textbf{15.17} \\
        Slim-UNETR \cite{pang2023slim} &83.75 / \textbf{87.08} &55.09 / \textbf{61.51} &41.94 / \textbf{48.29} &87.17 / \textbf{88.30} &54.59 / \textbf{58.21} &46.87 / \textbf{51.18} &55.79 / \textbf{56.10} &90.72 / \textbf{91.78} &61.83 / \textbf{62.86} &50.52 / \textbf{52.97} &41.47 / \textbf{48.22} \\
        Swin-UNETR \cite{hatamizadeh2021swin} &88.80 / \textbf{89.26} &61.84 / \textbf{64.06} &49.36 / \textbf{53.92} &90.12 / 89.93 &59.61 / \textbf{62.81} &47.92 / \textbf{51.62} &52.71 / \textbf{55.09} &93.53 / 93.00 &64.65 / 64.42 &45.93 / \textbf{51.58} &44.67 / \textbf{50.28} \\
        nnFormer \cite{zhou2023nnformer} &86.75 / \textbf{87.63} &57.41 / \textbf{62.31} &49.82 / \textbf{52.12} &89.12 / 88.88 &56.29 / \textbf{59.46} &50.29 / \textbf{52.75} &58.51 / 57.32 &92.48 / 91.93 &63.82 / \textbf{64.54} &48.71 / \textbf{53.59} &46.31 / \textbf{52.26} \\

        VSmTrans \cite{liu2024vsmtrans} &86.72 / \textbf{90.51} & 56.35 / \textbf{63.58} & 51.28 / \textbf{54.67} & 89.38 / \textbf{90.21} & 57.35 / \textbf{58.90} & 48.73 / \textbf{54.80} & 51.48 / \textbf{58.02} &92.76 / \textbf{93.61} & 61.72 / \textbf{63.01} & 47.84 / \textbf{51.99} & 43.73 / \textbf{50.38} \\
        \rowcolor[gray]{0.9} % 设为浅灰色
        Ours  &\textcolor{red}{93.35} &\textcolor{red}{73.31} &\textcolor{red}{59.33} &\textcolor{red}{91.75} &\textcolor{red}{68.61} &\textcolor{red}{56.06} &\textcolor{red}{62.08} &\textcolor{red}{94.90} &\textcolor{red}{73.38} &\textcolor{red}{57.06} &\textcolor{red}{58.65} \\
        \hline 
        \specialrule{0em}{0.2pt}{0.2pt}
        \hline
      \end{tabular}
       }
\end{table*}

\subsubsection{Assessment of and robustness}
Our method exhibited strong robustness, consistently delivering high segmentation performance across the AMOS, BraTS, and PENGWIN datasets, despite substantial differences in data characteristics, anatomical structures, and sample distributions.
Fig. \ref{fig: gain-amos-pengwin-dsc} (left) shows an inverse correlation between anatomical structure volume and DSC improvement on the AMOS dataset, with the smallest organs (LGA and RGA) achieving over 10\% improvement, highlighting the effectiveness of our contour-weighted strategy for handling volume disparities.
A similar trend is observed on the PENGWIN dataset (Fig. \ref{fig: gain-amos-pengwin-dsc}, right), where fragments with volume fractions below 10\% exhibit DSC gains exceeding 2\%, and even fragments F3 and F14, with only 7 and 9 samples respectively, achieve over 6\% improvement, demonstrating CWCD’s ability to address both volume and sample imbalance.

\subsubsection{Assessment of and model-independence}
\begin{figure}[t]
    \centering
    \includegraphics[width=\linewidth]{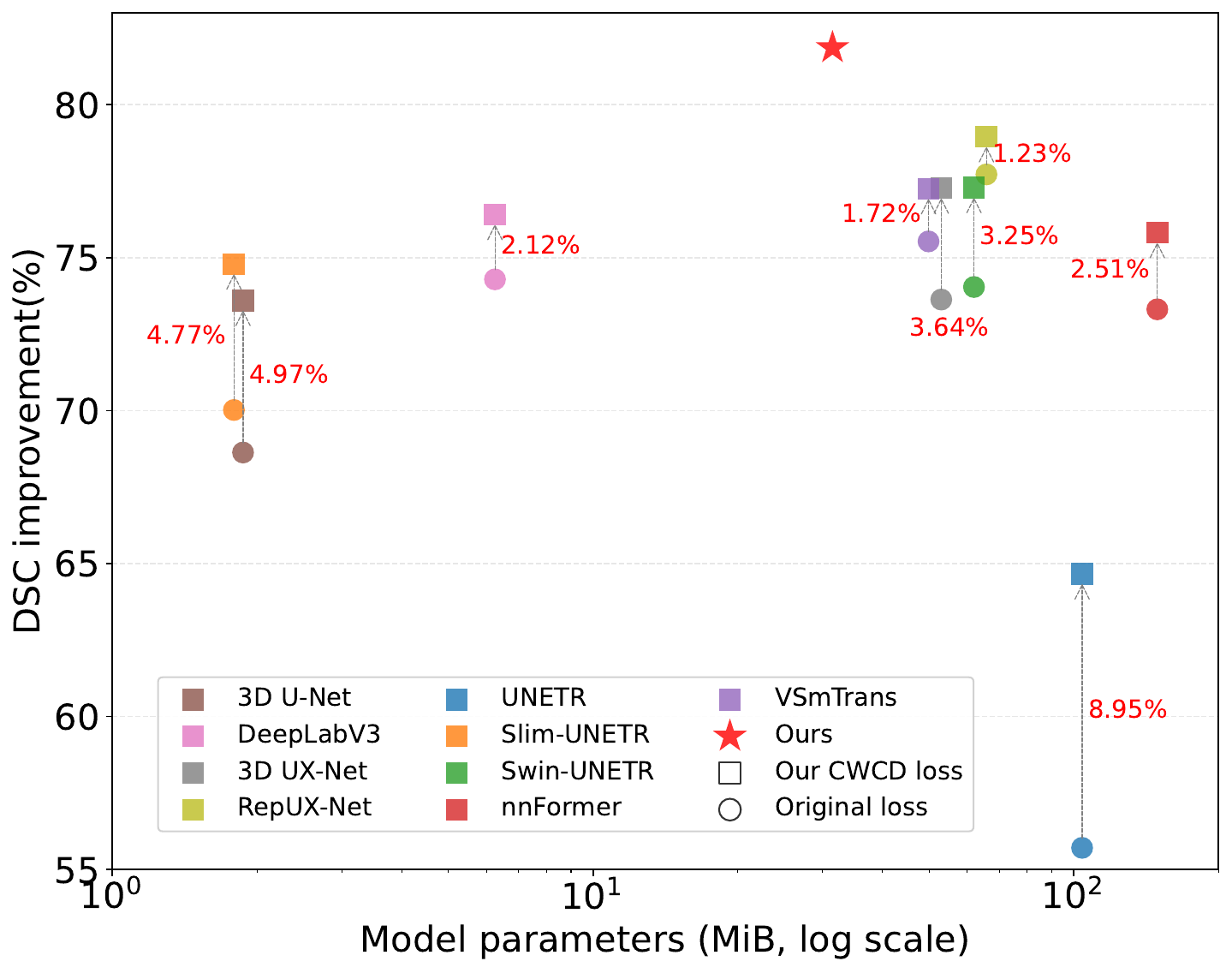}
    \caption{Segmentation performance (DSC improvement) versus model size for different comparison methods using our CWCD loss versus original losses on the PENGWIN dataset.}
    \label{fig: DSC-models-pengwin}
\end{figure}

Our method demonstrated flexibility and model independence, as it was seamlessly integrated into various segmentation frameworks to improve performance.
As shown in Table \ref{tab: different models}, on the AMOS and BraTS datasets, both lightweight (3D U-Net: +4.64\% / +3.05\% DSC) and computationally intensive architectures (UNETR: +2.56\% / +2.9\% DSC) showed improvements when equipped with our CWCD loss. The benefits were even more pronounced on the PENGWIN dataset, where 3D U-Net achieves remarkable gains (DSC: +4.97\%, HD95: +16.18\%, ASSD: +26.86\%) and UNETR also demonstrated prominent improvements (DSC: +8.95\%, HD95: +23.21\%, ASSD: +22.77\%). 
Notably, this contour-weighted strategy not only enhanced performance for these two representative architectures but also consistently improved segmentation results across nine comparison methods, yielding average Dice score improvements of 2.39\%, 1.44\%, and 3.88\% on AMOS, BraTS, and PENGWIN datasets, respectively. Fig. \ref{fig: DSC-models-pengwin} illustrates the improvements achieved by different models of varying parameter size when equipped with our contour-weighted strategy.
Our experiments revealed that simpler models, with our loss function applied, could rival or outperform more complex architectures in segmentation performance. For instance, On the PENGWIN dataset, the lightweight 3D U-Net (1.87M) with CWCD loss not only matched the original 3D UX-Net's performance but also outperformed nnFormer (149.17M) ($73.60\% ~vs~ 73.31\%$). This suggests that our method can effectively compensate for the limited feature representation capacity of lightweight models. 
These consistent improvements across architectures of varying complexity validated the flexibility and model independence of our method, highlighting its potential as a universal enhancement to existing segmentation frameworks.
Despite having fewer parameters than most competing models, PDANet (31.48M) consistently achieved superior DSC scores across all three datasets. This success validated that carefully designed loss functions can enable relatively simple architectures to achieve state-of-the-art performance, offering a more efficient alternative to address data imbalance.

% FIG. 
% =======

\section{Conclusion}
\label{sec: Conclusion}
In this paper, we have proposed a novel, model-independent approach that incorporates a contour-weighted strategy to effectively address data imbalance issues in medical image segmentation. We have also designed a lightweight network architecture, called PDANet, which integrates the partial decoder mechanism.
The proposed method has improved the network's ability to capture fine-grained details along the boundaries of segmentation targets by assigning weights to the cross-entropy loss based on contour maps and introducing a separable dice loss. Extensive experiments on three challenging datasets have demonstrated that our approach consistently outperformed existing methods, achieving average Dice score improvements of 2.39\%, 1.44\%, and 3.88\% on AMOS, BraTS, and PENGWIN datasets, respectively, when applied to nine state-of-the-art segmentation architectures.
The proposed method is model-independent and can be easily integrated into various segmentation frameworks to enhance their accuracy, from lightweight networks to computationally intensive frameworks. 
Notably, our CWCD loss enables lightweight models to achieve performance comparable to or exceeding that of more complex networks.
This flexibility highlights its practical value and potential for widespread applicability in addressing data imbalance challenges.

Although our method demonstrates strong performance across multiple datasets, several limitations remain. Currently, a uniform parameter setting is used for contour extraction across all segmentation targets, which does not fully accommodate variations in anatomical shape and volume. In future work, we plan to refine the contour extraction process by introducing adaptive strategies tailored to individual structures. Furthermore, we aim to enhance the flexibility and adaptability of the loss function to further improve segmentation performance.

\bibliographystyle{IEEEtran}
\bibliography{refs}
\end{document}